\documentclass{emulateapj}
\usepackage{graphicx}
\shorttitle{Spatially Compact Ly$\alpha$ emission in LAEs at z=3.1}
\shortauthors{Bond et al.}

\begin{document}

\title{EVIDENCE FOR SPATIALLY COMPACT L\lowercase{y}$\alpha$ EMISSION IN \lowercase{$z$}~$=3.1$ 
L\lowercase{y}$\alpha$-EMITTING GALAXIES\footnote{Based on observations made with the NASA/ESA Hubble Space Telescope, obtained from the data archive 
at the Space Telescope Institute. STScI is operated by the association 
of Universities for Research in Astronomy, Inc. under the NASA contract NAS 5-26555.}}

\author{Nicholas A. Bond}
\affil{Physics and Astronomy Department, Rutgers University,
    Piscataway, NJ 08854-8019, U.S.A.}
\email{nbond@physics.rutgers.edu}
\author{John J. Feldmeier}
\affil{Department of Physics and Astronomy, Youngstown State University, Youngstown, OH 44555, U.S.A.}
\email{jjfeldmeier@ysu.edu}
\author{Ana Matkovi\'{c}, Caryl Gronwall, Robin Ciardullo}
\affil{Department of Astronomy and Astrophysics, Pennsylvania State University, University Park, PA 16802, U.S.A.} 
\email{matkovic@astro.psu.edu, caryl@astro.psu.edu,rbc@astro.psu.edu}
\and 
\author{Eric Gawiser}
\affil{Physics and Astronomy Department, Rutgers University,
    Piscataway, NJ 08854-8019, U.S.A.}
\email{gawiser@physics.rutgers.edu}

\begin{abstract} We present the results of a high-spatial-resolution study of the line emission in a sample of $z=3.1$ Ly$\alpha$-Emitting Galaxies (LAEs) in the Extended Chandra Deep Field-South.  Of the eight objects with coverage in our {\it HST}/WFPC2 narrow-band imaging, two have clear detections and an additional two are barely detected ($\sim 2 \, \sigma$).  The clear detections are within $\sim 0.5$~kpc of the centroid of the corresponding rest-UV continuum source, suggesting that the line-emitting gas and young stars in LAEs are spatially coincident.  The brightest object exhibits extended emission with a half-light radius of $\sim 1.5$~kpc, but a stack of the remaining LAE surface brightness profiles is consistent with the WFPC2 point spread function.  This suggests that the Ly$\alpha$ emission in these objects originates from a compact ($\lesssim 2$~kpc) region and cannot be significantly more extended than the far-UV continuum emission ($\lesssim 1$~kpc).  Comparing our WFPC2 photometry to previous ground-based measurements of their monochromatic fluxes, we find at $95$\% ($99.7$\%) confidence that we cannot be missing more than $22$\% ($32$\%) of the Ly$\alpha$ emission.  \end{abstract}

\keywords{cosmology: observations --- galaxies: formation --- galaxies: high-redshift --- galaxies: structure}

\vspace{0.4in}

\section{INTRODUCTION}

Lyman-Alpha-Emitting (LAE) galaxies at $z \sim 2 - 4$ are thought to be actively star-forming \citep[e.g.][]{CowieHu}, with low stellar masses ($\sim 10^9 M_{\sun}$), high mass-specific star-formation rates, and very little dust \citep{Venemans05,Gawiser07}.  Morphological studies of their rest-frame ultraviolet light using the {\it Hubble Space Telescope} ({\it HST}) reveal small ($r_e\lesssim 1$~kpc), compact ($C>2.5$), galaxies that are often clumpy or irregular \citep{Venemans05,Pirzkal07,Overzier08,Taniguchi09,BondLAE,GronwallMorph}.


Although the objects known as Lyman Alpha blobs often exhibit extended diffuse morphologies \citep[e.g.][]{FirstBlobs,Nilsson06,SJ07} that could be explained by cold accretion onto a dark matter halo \citep{Dijkstra06} or AGN/stellar feedback \citep{Geach09}, to date there has been no published study of the morphology of the line-emitting regions of the lower-mass LAEs.  In the local universe, {\it HST} observations by \citet{Ostlin08} reveal that much of the Ly$\alpha$ emission of six star-forming galaxies originates in a diffuse halo surrounding the galaxy.  These ``Ly$\alpha$ halos'' are likely caused by resonant scattering of the Ly$\alpha$ line off of cold, diffuse gas surrounding the galaxy.  If the same process occurs at high redshift, then we would expect LAEs to exhibit extended Ly$\alpha$ emission and, since the majority of LAEs are resolved in the rest-UV continuum at {\it HST} resolution \citep{BondLAE}, the extended Ly$\alpha$ halos should be detectable.  In a study of LAEs at $z=2.25$, \citet{NilssonLAE} report that the majority of their sample are ``significantly more extended than point sources'' in narrow-band images with $1$\arcsec\ seeing.  If true, this would suggest an order-of-magnitude difference between the extent of the UV continuum emission and the line emission.

The Multiwavelength Survey by Yale-Chile \citep[MUSYC,][]{MUSYC} is a broad-based collaboration dedicated to obtaining multiwavelength imaging and spectroscopy of $1.2$~degree$^2$ of sky and includes coverage of the Great Observatories Origins Deep Survey South (GOODS-S) region within the Extended Chandra Deep Field-South (ECDF-S).  As part of this survey, \citet{GronwallLAE} discovered an unbiased sample of 162 LAEs at $z=3.1$ using broadband and 4990~\AA\ narrow-band imaging.  In a previous paper, we presented an analysis of the rest-frame ultraviolet light of $97$ of these LAEs \citep{BondLAE} and found that the majority were small ($\lesssim 1$~kpc), single-component objects.  \citet{GronwallMorph} extend this analysis to higher-order morphological diagnostics, finding that LAEs possess a wide range of S\'{e}rsic indices and ellipticities, with $\langle\epsilon\rangle\simeq 0.5$.


In this paper, we present the results of a high-spatial-resolution study of the line emission in 8 LAEs drawn from the \citet{GronwallLAE} sample.  Throughout we will assume a concordance cosmology with $H_0=71$~km~s$^{-1}$~Mpc$^{-1}$, $\Omega_{\rm m}=0.27$, and $\Omega_{\Lambda}=0.73$ \citep{WMAP}.  With these values, $1\arcsec = 7.75$~physical~kpc at $z=3.1$.

\section{DATA}
\label{sec:data}

\begin{figure}
\epsscale{1.0}
\plotone{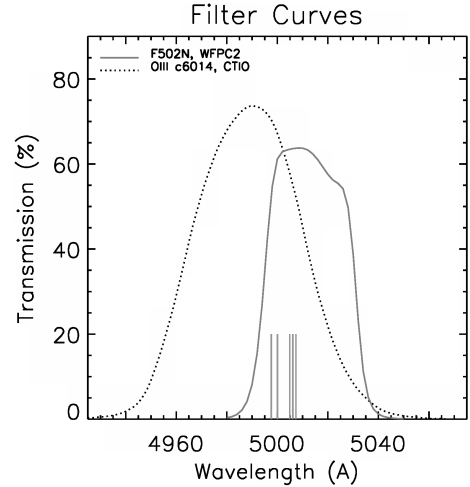}
\caption{Filter transmission curves for the CTIO NB4990 filter (c6014, dotted curve) and the HST/WFPC2 F502N filter (solid curve).  The vertical solid lines at the bottom of the plot indicate the wavelengths of the Ly$\alpha$ lines in each of the spectroscopically-confirmed LAEs (see Table~\ref{tab:PhotTable})  Note that LAE 16 and LAE 117 have the same redshift (to the quoted precision) and are therefore only visible as one line on the plot.
\label{fig:FilterCurves}}
\end{figure}

Minor modifications to the \citet{GronwallLAE} sample were published in \citet{BondLAE}.  In both of these studies, LAEs were selected to have $F > 1.5 \times 10^{-17}$~ergs~cm$^{-2}$~s$^{-1}$ in a $4990$\AA~narrow-band filter, and an observed-frame Ly$\alpha$ equivalent width $>80$~\AA\null.  As shown in Figure~\ref{fig:FilterCurves}, the filter used to create the \citet{GronwallLAE} sample only overlaps with the {\it HST} narrow-band filter over a $\sim 20$~\AA\ wavelength range, so our two {\it HST}/WFPC2 pointings targeted LAEs with Ly$\alpha$ lines identified spectroscopically within this region (which corresponds to $3.11 \lesssim z\lesssim 3.13$, see Table~\ref{tab:PhotTable}).  The fields were observed as program GO-11177 (PI: C. Gronwall) on July 28-30 and August 13-14 2008.  Each was observed with twelve exposures of $2600$~s each through the F502N filter applying a standard four-point dither pattern.  Once the data were taken, we generated stacked images and weight images from our pipeline-reduced exposures using the \textit{MultiDrizzle} task within PyRAF.  We found that the parameters PIXFRAC$=0.8$ and PIXSCALE$=0.05$\arcsec\ optimized a combination of image quality and cosmic-ray rejection efficiency.  Other input parameters to \textit{MultiDrizzle} were set to their standard values \citep[see ][]{FH02,FS09}.

The two {\it HST}/WFPC2 pointings lie within the southern GOODS field, which covers $\sim 160$~arcmin$^2$ of sky and includes {\it HST}/ACS observations in the $B_{435}$, $V_{606}$, $I_{775}$, and $z_{850}$ filters.  In \citet{BondLAE}, the $V_{606}$-band images were used to identify the rest-frame ultraviolet centroids and sizes of each LAE component, so we use these measurements for comparison to the emission-line centroids and morphologies measured with {\it HST}/WFPC2.  All GOODS broadband images were multidrizzled to a pixel scale of $0\farcs03$ pixel$^{-1}$ and the typical $5 \,\sigma$ detection limit in $V_{606}$ is $m_{\rm AB} = 28.8$.

In order to make a robust comparison between the centroids of the rest-frame ultraviolet continuum emission and the Ly$\alpha$ emission, we must ensure that the world coordinate system of the {\it HST}/WFPC2 images is matched to that of the GOODS $V_{606}$-band images.  Using a sample of five stars in field 1, we estimate $\Delta \alpha_0=+0\farcs19$ and $\Delta \delta_0=+0\farcs18$.  For field 2, we found $\Delta \alpha_0=-0\farcs15$ and $\Delta \delta_0=-0\farcs11$.  Each of these corrections is uncertain by $\sim 0\farcs03$ and individual stars were found to deviate by as much as $0\farcs05$ in a single direction.  

A single photometric calibration was performed for both WFPC2 images using a sample of three bright stars with well-behaved curves-of-growth.  We performed $1$\arcsec-aperture photometry on each star and calibrated their magnitudes to the MUSYC NB4990 image of the ECDF-S \citep{Gawiser06}.  Using the mean calibration to these stars, we found an AB zeropoint of $m_{AB}=19.87$ and a scatter of $0.05$~mag.  

\section{METHODOLOGY}
\label{sec:method}

\begin{figure}
\epsscale{1.0}
\plotone{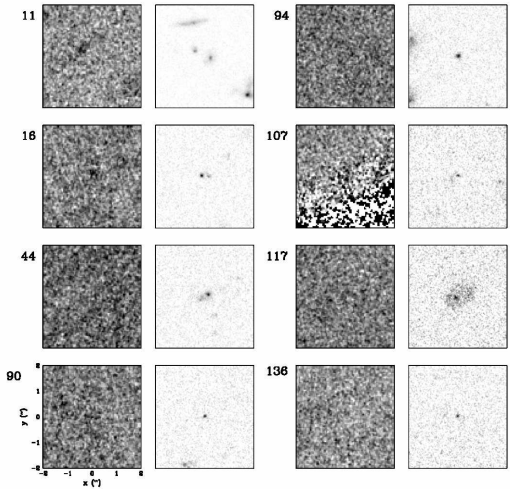}
\caption{LAE cutouts extracted from the WFPC2 F502N (left) and GOODS V-band (right) image for the four brightest LAEs.  All cutouts are $4$\arcsec\ ($31$~kpc) on a side and centered at the V-band centroid.  The panels are marked with their indices in the \citet{GronwallLAE} catalog.  Note that the indexing scheme places the LAEs in order of decreasing narrow-band flux as determined from the MUSYC ground-based images.
\label{fig:Cutouts}} 
\end{figure}

We extracted $8$\arcsec\ $\times 8$\arcsec\ pixel ($62 \times 62$~kpc) cutouts from the {\it HST}/WFPC2 images (see the left panels of Figure~\ref{fig:Cutouts}), each centered at the GOODS $V_{606}$-band centroid of the LAE.  Of the eight objects with coverage, two (LAEs 11 and 16) have clear detections near the cutout centers.  There is evidence for a possible detection in LAE 107, but it lies near the edge of a WFPC2 pointing and vignetting makes this difficult to verify.  We exclude this LAE from subsequent analysis due to poor data quality.

Following \citet{BondLAE}, we used SExtractor \citep{SExtractor} to identify
components within the WFPC2 images, fitting a constant sky to the cutout and defining detections as fifteen contiguous pixels above a $1.65 \, \sigma$ threshold.  This procedure confirmed single detections in LAEs 11 and 16, but no others within $0.5$\arcsec\ of the $V_{606}$-band LAE centroids.

We used the {\tt IRAF} routine PHOT to perform aperture photometry on each object in our sample, in each case using the minimum aperture that enclosed the total emission line flux seen within $0\farcs5$.  The mean aperture size used for the sample was $0\farcs34$.  For the two LAEs with clear detections, the aperture center was set to their SExtractor position, while the centers for the remaining LAEs were set to their $V_{606}$-band centroids \citep[see ][and the right panels of Figure~\ref{fig:Cutouts}]{BondLAE}.  Following \citet{GronwallLAE}, we obtain monochromatic fluxes,
\begin{equation}
F_{{\rm Ly}\alpha}=3.63 \times 10^{-20}\,10^{-m_{AB}/2.5}\frac{c}{\lambda^2}\frac{\int T(\lambda)d\lambda}{\left<T_{{\rm Ly}\alpha}\right>},
\end{equation}
where $F_{{\rm Ly}\alpha}$ is in erg~cm$^{-2}$~s$^{-1}$, $T(\lambda)$ is the filter transmission function, and $\left<T_{Ly\alpha}\right>$ is the mean value of the transmission function when convolved with the Ly$\alpha$ line profile \citep{Jacoby87}.  When computing $\left<T_{Ly\alpha}\right>$, we have approximated the Ly$\alpha$ line as a Gaussian with $FWHM=500$~km~s$^{-1}$ and centered at its spectroscopic redshift.  The resulting monochromatic fluxes are given in column~5 of Table~\ref{tab:PhotTable}.  As expected, LAEs 11 and 16 are $>2 \, \sigma$-detected using {\tt PHOT}.  LAEs 90 and 94 also prove to be $>2 \, \sigma$ detections. In all LAEs, the rest-UV continuum will make a small contribution to the total narrow-band flux, but its impact on the observed morphology is negligible.  Of the LAEs with {\tt PHOT} detections, the object with the largest relative continuum contribution (LAE 11) has an expected flux that is $7.5$ times smaller than the measured narrow-band flux. 

We also report the half-light radius of each source, interpolated individually from the curves-of-growth.  For the LAEs without SExtractor detections, these half-light radii will be biased high due to uncertainty in the source centers \citep[see ][]{BondLAE}.  

\section{RESULTS}
\label{sec:results}

The offsets between the emission-line and rest-UV-continuum centroids of the two LAEs with SExtractor detections are given in column~6 of Table~\ref{tab:PhotTable}.  In LAE 11, the emission-line centroid is clearly offest from the rest-UV centroid ($\sim 10 \, \sigma$), but is $0\farcs04$ from the position of the northwest component in the V-band image (see Figure~\ref{fig:Cutouts}), suggesting that this component is the host for the Ly$\alpha$ emission.  The measured Ly$\alpha$ centroid of LAE16 is offset from the rest-UV centroid by $0\farcs06$, consistent with the expected uncertainties in the astrometric correction and source centroiding (see \S~\ref{sec:data}).  

We also computed surface brightness profiles for sets of objects in the F502N images.  To do this, we first performed photometry on individual objects within a series of apertures, increasing in increments of $0.75$~pixels ($0\farcs0375$).  The resulting curves-of-growth were each normalized to their total flux and averaged, weighting by the inverse variance in each bin.  The final stacked surface brightness profiles were computed from this stacked curve-of-growth.

\begin{figure}
\epsscale{1.0}
\plotone{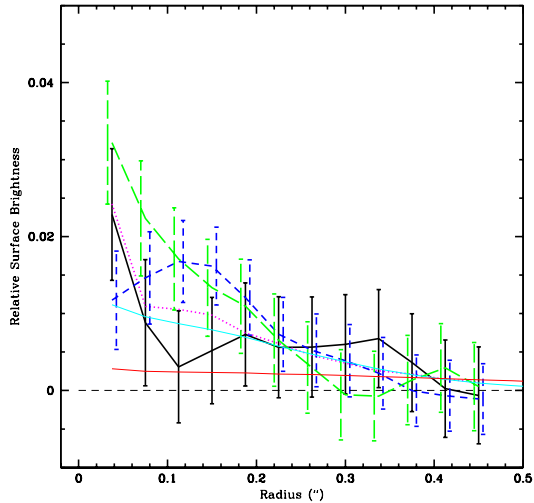}
\caption{Normalized F502N surface brightness profiles for LAE 11 (short dashed curve), LAE 16 (dot-dashed curve), a stack of the other five LAEs (solid curve), and five stars in the WFPC2 images (dotted curve).  In addition, we plot the surface brightness profiles for Gaussian models with $\sigma=0\farcs2$ ($1.5$~kpc) and $\sigma=0\farcs4$ ($3$~kpc) as thin solid lines.  Each curve is normalized to the total light from the source.
\label{fig:SBProfile}} 
\end{figure}

The results of this procedure are plotted in Figure~\ref{fig:SBProfile}, which shows the normalized F502N surface brightness profiles for LAE 11, LAE 16, a stack of five stars in the \citet{MartinStars} catalog, and a stack of all LAEs in the sample except for LAE 11, LAE 16, and LAE 107.  In addition, we generated a series of two-dimensional Gaussian models for each source, normalizing each model to its total flux.   In Figure~\ref{fig:SBProfile}, we overplot models with $\sigma=0\farcs2$ and $0\farcs4$ (neither convolved with the PSF).  The stack of the LAE surface brightness profiles is consistent with that of stars and is marginally consistent with the $\sigma=0\farcs2$ model.  The surface brightness profile of LAE 16 is consistent with that of stars and does not fit a model that is much more extended than the $\sigma\simeq 0\farcs12$ PSF.  The $\sigma=0\farcs4$ model is a poor fit to all of the plotted surface brightness profiles.  This suggests that the line-emitting region in the majority of LAEs is compact ($\lesssim 2$~kpc) and cannot be significantly more extended than the far-UV continuum emission ($\lesssim 1$~kpc).  

The exception is LAE 11, which exhibits extended line emission out to $\sim 0\farcs2$ ($\sim 1.5$~kpc) from the source center.  Although the $V$-band (UV-continuum) image suggests the presence of at least two components in the system, the line emission is centered on the northwest component and their separation is too great ($\sim 0\farcs7$) for the second object to be the dominant source of the extended emission.  The narrow-band image also shows evidence for two distinct clumps of emission, but at a much smaller separation ($\sim 0\farcs15$).  This could indicate the presence of multiple star-forming regions, a dust lane, or perhaps Ly$\alpha$ photons from a single star-forming region scattering from multiple clumps of gas.  Even considering this substructure, however, the line-emitting region is still compact ($\lesssim 1.5$~kpc) and should be indistinguishable from a point source in ground-based imaging.

Analysis of the surface brightness profiles can miss an extended diffuse component if the mean surface brightness is below the sky noise.  To explore this possibility, we compare the monochromatic fluxes of our LAEs as measured in the WFPC2 frames to those measured from the ground by \citet[][see column~6 of Table~\ref{tab:PhotTable}]{GronwallLAE}.  In all cases, the fluxes measured from space are within $1.5 \, \sigma$ of those measured from the ground.  We cannot rule out the presence of extended emission in the individual undetected LAEs in our sample, but we find that a model in which we are missing $>22$\% ($32$\%) of the line emission within $0\farcs5$ of each LAE is ruled at $>95$\% ($99.7$\%) confidence.

\section{DISCUSSION}
\label{sec:discussion}

The results presented in \S~\ref{sec:results} suggest that the Ly$\alpha$ emission in the majority of LAEs originates from a small ($\lesssim 2$~kpc) region that coincides very closely (within $\sim 0.5$~kpc) with the young stars producing the rest-frame UV continuum emission.  One object in our sample (LAE 11) exhibits extended line emission out to $\sim 1.5$~kpc, but this is still comparable in extent to the rest-UV continuum \citep[$r_e \simeq 1.3$~kpc,][]{BondLAE}.  

Unlike the LAEs in our sample, low-redshift star-forming galaxies typically exhibit extended Ly$\alpha$ halos \citep{Ostlin08} out to many half-light radii from the rest-frame UV emission.  These galaxies are not direct analogs of the objects in our sample, as both their Ly$\alpha$ luminosities and specific star formation rates are far smaller than those of the typical LAE.  However, their Ly$\alpha$ halos suggest the presence of a reservoir of gas at large radii that is either not present in $z=3.1$ LAEs or is only scattering a small fraction of the Ly$\alpha$ photons created by star formation.

Although this paper represents the first published search for extended Ly$\alpha$ emission in LAEs using {\it HST} imaging, \citet{NilssonLAE} report that $z=2.25$ LAEs are ``significantly more extended than point sources'' in narrow-band images with $1$\arcsec\ seeing.  There may be a modest amount of size evolution in LAEs between $z=3.1$ and $z=2.25$, but their result would seem to contradict our finding that even the most extended LAEs have emission-line half-light radii $<0.2$\arcsec.  It is possible that we have failed to account for a component of highly diffuse line emission extending to large radii, but we find it must be $<22$\% of the total line emission in the present sample.

It is important that we continue to explore the wavelength-dependence of high-redshift galaxy morphologies, as different regions of the spectrum probe different physical components of the host galaxy.  The sample presented here could certainly be improved, both with greater sky coverage and deeper observations.  It would also be interesting to target a sample of Ly$\alpha$-emitting Lyman Break Galaxies to determine how the spatial distribution of their Ly$\alpha$ emission differs from that of a typical LAE, if at all.  

\acknowledgments

We would like to thank Patrick Durrell for helpful discussions about {\it HST}/WFPC2 data reduction.  We would also like to thank the referee for their helpful comments.  Support for this work was provided by NASA through grant numbers HST-GO-11177.01, HST-AR-11253.01-A, and HST-AR-10324.01 from the Space Telescope Science Institute, which is operated by AURA, Inc., under NASA contract NAS 5-26555 and by the National Science Foundation under grants AST-0807570 and NSF AST-0807885.

\bibliographystyle{apj}                       




\begin{deluxetable}{lccccccc}
\tablecaption{LAE Photometric Properties in Narrow-Band WFPC2 Images\label{tab:PhotTable}}
\tablewidth{0pt}
\tablehead{
\colhead{ID\tablenotemark{a}}
&\colhead{$\alpha$\tablenotemark{b}}
&\colhead{$\delta$\tablenotemark{b}}
&\colhead{$z$}
&\colhead{${\rm log}$~$F_{{\rm Ly}\alpha}$}
&\colhead{${\rm log}$~$F_{{\rm Ly}\alpha}^{{\rm G07}}$}
&\colhead{$d_c$\tablenotemark{c}}
&\colhead{$r_e^{\rm PHOT}$~\tablenotemark{d}}
\\
&
&
&
&\colhead{(erg/s/cm$^2$)}
&\colhead{(erg/s/cm$^2$)}
&\colhead{(\arcsec)}
&\colhead{(\arcsec)}
}
\startdata
$11$ &$3$:$32$:$26.922$ &$-27$:$41$:$28.174$ &$3.113$ &$-16.03 \pm 0.12$ &$-16.11 \pm 0.02$ &$0.51$ &$0.17$  \\
$16$ &$3$:$32$:$13.274$ &$-27$:$43$:$29.933$ &$3.117$ &$-16.28 \pm 0.14$ &$-16.22 \pm 0.03$ &$0.06$ &$0.14$  \\
$44$ &$3$:$32$:$15.783$ &$-27$:$44$:$10.228$ &$3.119$ &$-17.32 \pm 1.52$ &$-16.43 \pm 0.04$ &$...$ &$0.22$  \\
$90$ &$3$:$32$:$14.560$ &$-27$:$45$:$52.634$ &$3.118$ &$-16.34 \pm 0.22$ &$-16.64 \pm 0.05$ &$...$ &$0.24$  \\
$94$ &$3$:$32$:$09.321$ &$-27$:$43$:$54.427$ &$3.113$ &$-16.41 \pm 0.29$ &$-16.65 \pm 0.05$ &$...$ &$0.29$  \\
$107$ &$3$:$32$:$10.143$ &$-27$:$44$:$28.558$ &$...$ &$...$ &$-16.68 \pm 0.06$ &$...$ &$...$  \\
$117$ &$3$:$32$:$12.936$ &$-27$:$44$:$51.590$ &$3.117$ &$-16.58 \pm 0.35$ &$-16.70 \pm 0.06$ &$...$ &$0.18$  \\
$136$ &$3$:$32$:$24.315$ &$-27$:$41$:$52.125$ &$3.111$ &$-17.13 \pm 1.54$ &$-16.76 \pm 0.07$ &$...$ &$0.24$  \\
\enddata
\tablenotetext{a}{Index from table 2 of Gronwall et al. 2007}
\tablenotetext{b}{Position of WFPC2 narrow-band centroid (set to the V-band centroid when there are no SExtractor detections)}
\tablenotetext{c}{Distance between WFPC2 and ACS V-band centroids, LAE 11 is $0\farcs04$ from the position of the northwest component in the V-band image}
\tablenotetext{d}{Half-light radius computed by {\tt PHOT}}
\end{deluxetable}

\end{document}